\documentclass[a4paper,10pt,DIV=calc]{scrartcl}

\usepackage{cancel}
\usepackage{tabularx}
\usepackage{subfig}
\usepackage{a4}
\usepackage{amsmath,amssymb}
\usepackage{bbm}
\usepackage{amsthm}
\usepackage{dsfont}
\usepackage[onehalfspacing]{setspace} 
\usepackage{appendix}
\usepackage{graphicx}
\usepackage{enumerate}
\usepackage{hyphenat}
\usepackage{cite}
\usepackage{colonequals}
\usepackage{relsize}
\usepackage{caption}
\usepackage{setspace}
\usepackage{MnSymbol}

\usepackage{enumerate}
\usepackage{relsize}

\usepackage[usenames,dvipsnames]{xcolor}
\usepackage[all]{xypic}

\usepackage{accents}
\usepackage{amsmath,amssymb}

 \theoremstyle{plain} 
\theoremstyle{plain} 
\theoremstyle{plain} 
\theoremstyle{plain} 
\theoremstyle{plain} 
\theoremstyle{definition} \newtheorem{Def}{Definition}
\theoremstyle{plain} 
\theoremstyle{definition}
\theoremstyle{definition} \newtheorem{Example}{Example}

\makeatletter
\newsavebox\myboxA
\newsavebox\myboxB
\newlength\mylenA

\newcommand*\xo[2][0.75]{%
    \sbox{\myboxA}{$\m@th#2$}%
    \setbox\myboxB\null
    \ht\myboxB=\ht\myboxA%
    \dp\myboxB=\dp\myboxA%
    \wd\myboxB=#1\wd\myboxA
    \sbox\myboxB{$\m@th\overline{\copy\myboxB}$}
    \setlength\mylenA{\the\wd\myboxA}
    \addtolength\mylenA{-\the\wd\myboxB}%
    \ifdim\wd\myboxB<\wd\myboxA%
       \rlap{\hskip 0.5\mylenA\usebox\myboxB}{\usebox\myboxA}%
    \else
        \hskip -0.5\mylenA\rlap{\usebox\myboxA}{\hskip 0.5\mylenA\usebox\myboxB}%
    \fi}
\makeatother

\usepackage{bbm}
\usepackage{amsthm}
\usepackage{appendix}
\usepackage{graphicx}
\usepackage[para]{footmisc}
\usepackage{tablefootnote}

\usepackage[para]{threeparttable}
\usepackage{etoolbox}
\usepackage{lipsum}
\usepackage[all]{xypic}
\usepackage{color}
\AtEndEnvironment{threeparttable}{\vskip10pt}{}{}
\makeatletter
\patchcmd{\TPT@doparanotes}
  {\TPT@hsize}
  {\TPT@hsize\footnotesize}
  {}
  {}
\makeatother

\graphicspath {{figures/}}

\definecolor{mygray}{gray}{0.7}

\DeclareOldFontCommand{\bf}{\normalfont\bfseries}{\mathbf}
\usepackage{amsmath}
\usepackage{amsfonts}
\usepackage{amssymb}

\newcommand{\pef}{$f_{\mathrm{\smaller PE}}$\,}
\newcommand{\pefm}{f_{\mathrm{\smaller PE}}}

  \usepackage{authblk}
\usepackage{fancyhdr}

\usepackage{hyperref}
\usepackage{cleveref}
 \usepackage{tikz}
\usetikzlibrary{shapes,arrows}
\usetikzlibrary{snakes}

\makeatletter
\def\@maketitle{%
  \newpage
  \null
  \vskip 2em%
  \begin{center}%
  \let \footnote \thanks
    {\Large\bf  \@title \par}%
    \vskip 1.5em%
    {\normalsize
      \lineskip .5em%
      \begin{tabular}[t]{c}%
        \@author
      \end{tabular}\par}%
    \vskip 1em%
    {\normalsize \@date}%
  \end{center}%
  \par
  \vskip 1.5em}
\makeatother
\title{Sentiment Protocol: A Decentralized Protocol Leveraging Crowd Sourced Wisdom}

\author{Anton Muehlemann%
\thanks{\texttt{muehle@berkeley.edu}}}
\affil{\small\textit{University of California, Berkeley} \\ \textit{94720, Berkeley,USA}}

\date{Dated: \today}

 \hypersetup{
  pdfauthor = {Anton Muehlemann (University of California, Berkeley)},
  pdftitle = {Sentiment protocol - a decentralized protocol leveraging crowd sourced wisdom},
  pdfsubject = {Sentiment protocol},
  pdfkeywords = {sentiment, crowd-sourced wisdom, surveys, decentralized, smart-contract, ethereum, EVM, blockchain}
}

\tikzstyle{block} = [draw, fill=blue!20, rectangle, 
    minimum height=3em, minimum width=6em]
\tikzstyle{sum} = [draw, fill=blue!20, circle, node distance=1cm]
\tikzstyle{input} = [coordinate]
\tikzstyle{output} = [coordinate]
\tikzstyle{pinstyle} = [pin edge={to-,thin,black}]
\newcommand{\dat}[3]{$#1$-$#2$-$#3$}

\renewcommand{\O}{\ensuremath{\mathcal O}}
\renewcommand{\S}{\ensuremath{\mathcal S }}

\newcommand{\tok}{\textsc{Token}}
\newcommand{\sen}{\textsc{Sentiment}}

\usepackage{pgfplots}

\usepackage{url}

\begin{document}

\thispagestyle{empty}

\pagenumbering{roman}
\thispagestyle{empty}
\clearpage
\maketitle
\begin{abstract}
The wisdom of the crowd is a valuable asset in today's society. It is not only important in predicting elections but also plays an essential role in marketing and the financial industry. Having a trustworthy source of opinion can make forecasts more accurate and markets predictable. Until now, a fundamental problem of surveys is the lack of incentives for participants to provide accurate information. Classical solutions like small monetary rewards or the chance of winning a prize are often not very attractive for participants. More attractive solutions, such as prediction markets, face the issue of illegality and are often unavailable. In this work, we present a solution that unites the advantages from classical polling and prediction markets via a customizable incentivization framework. Apart from predicting events, this framework can also be used to govern decentralized autonomous organizations.\\
 \vspace{4pt}

 \thispagestyle{empty}


\vspace{4pt}

\noindent\textsc{Tx id:}
\noindent\href{https://etherscan.io/tx/0x0173da4bbf1aafbcfbf75c38b07aeebe399c6f7dfe06f8b4065524d935f726b8}{\smaller \noindent \texttt{0x0173da4bbf1aafbcfbf75c38b07aeebe399c6f7dfe06f8b4065524d935f726b8}}
\end{abstract}
\clearpage
\newpage
\tableofcontents

\newpage
\clearpage
\cleardoublepage\pagenumbering{arabic}
\setcounter{page}{1}
\section{Introduction}
Sentiment in its broadest sense, is of great interest to both politics and industry. In the USA alone, the revenue of market research and public opinion polling reached $\$18$B in 2016 and continues to grow\cite{polling}. In spite of its size, political forecasts have failed to accurately predict two major recent political events, the US presidential elections and the decision of the UK to leave the EU\cite{president,brexit}. These failures may come as a surprise as, with the emergence of social networks and generally more interactive website, it has never been easier to source the opinion of the crowd. 

In the financial industry, rating agencies are highly-paid providers of sentiment on a wide range of investment vehicles. However, if one takes for example a popular US stock such as TESLA, on Oct 13, 2017 out of $25$ analysts, $35\%$ recommended \emph{buy}, another $35\%$ recommended \emph{hold} and the remaining $30\%$ recommended \emph{sell}\cite{tesla}. Therefore, making the prediction no better than a simple guess\cite{Monkey}. Another well document fact is that most actively managed funds fail to beat the market, showing again that so-called expert opinions are not as valuable as they may seem\cite{buffett,Mutual}. 

A common feature of the above cases is that repercussions (and rewards) for providing inaccurate (accurate) sentiment are extremely limited for both individuals, such as experts, and the crowd. For experts, this is not at last due to the difficulty of objectively evaluating their performance. Reasons include the phenomenon of survival bias, i.e. only the experts that made good predictions are cited, and also, if the expert is influential enough, the ability of turning ones prediction into a self-fulfilling prophecy. For crowd sentiment, some opinion polls offer small monetary rewards, usually in the form of a lottery, for filling out a survey. However, these rewards are paid irrespectively of the quality of the provided sentiment and in particular does not prevent anyone from intentionally providing false statements. Furthermore, specifically in web surveys, participants are able to repeatedly reply to the same surveys to increase their potential reward. 

A different form and an arguably significantly more successful approach of information aggregation are speculative markets\cite{Hirshleifer,Berg,Berg2008,Epley,Futarchy,VitalikFut,Augur,Gnosis}. There, free bidding markets of outcome shares are offered and, dictated by economic theory, share prices will become representative of the likelihood of an event coming true. Unfortunately, under most legislations (in particular in the US), speculative markets are considered gambling and thus illegal - making this option often unavailable.

In the present article, we introduce a framework that unites the legal benefits of classical polling with the predictive power of speculative markets by introducing pre-defined reward functions whose payouts are both fixed and performance based. The structure of this article is as follows: At first, we introduce the Sentiment Protocol in Section \ref{SecSentProtocols}. The section starts with a high level overview and is followed by a detailed description of each component. Each stage includes basic examples that illustrate the respective concepts. In Section \ref{SecInc}, we analyze the incentives for sentiment contributors and pollsters. We also discuss possible vulnerabilities and give suggestions on how they may be overcome. In the final Section \ref{SecUse}, we discuss in further detail two use cases for the Sentiment Protocol. 
 
\section{The sentiment protocol}\label{SecSentProtocols}
It is now widely accepted that blockchain technology, the underlying concept of Bitcoin\cite{bitcoin}, has the ability to disrupt a wide range of industries including financial services, technology, media and telecommunications\cite{Deloitte,Wattenhofer}. Arguably, one of the most revolutionary emerging concepts are so-called smarts contracts which are scripts executed on a world spanning super computer\cite{Ethereum}. The Sentiment Protocol leverages these new possibilities to on the one hand immutably record provided sentiment and on the other hand conduct monetary transactions, such as rewards or penalties, without the need of trusting a third party. 

The novelty of the protocol is that it leverages the predictive power of speculative markets while retaining the legality of classical polling. This goal is achieved by introducing a performance based reward function $\pefm$ (cf. \Cref{PerformanceEvaluation}), resulting in higher payouts for better predictions, with a reward pool provided by the pollster. Since the pollster takes a distinguished role, speculative risks are taken away from the sentiment contributors.

Whenever the performance based reward function $\pefm$ only takes non-negative values, participants can {only} earn rewards. Recalling that by definition gambling is ``the act of risking money, or anything of value, on the outcome of something involving chance''\cite{Gambling}, it is clear that in the absence of a risk of loss, the sentiment protocol cannot be considered gambling. If the performance based reward function can also attain negative values, participants can loose part of their invested stake, making it more similar to classical prediction markets.

\subsection*{Protocol overview}
\begin{figure}[ht]
\begin{center}
  \begin{tikzpicture}


   \draw[->] (0,0) -- (12.15,0) node[anchor=north] {$T$};
 
 \draw (0,.25) -- (0,-.25);
 \draw (5,.25) -- (5,-.25);
 \draw (6.5,.25) -- (6.5,-.25);
 \draw (10,.25) -- (10,-.25);
 
  \tikzstyle{period} = [rectangle, rounded corners, minimum width=5cm, minimum height=.1cm,text centered, draw=black, fill=gray!30]
 \tikzstyle{point} = [rectangle, rounded corners, minimum width=.1cm, minimum height=.1cm,text centered, draw=black, fill=gray!30]
 
 \node (start) {};
 \node (sentiment) [period, right of=start,node distance=2.5cm] {};
 \node (tally) [point, right of=start,node distance=6.5cm] {};
 \node (pe) [point, right of=start,node distance=10cm] {};

 \tikzstyle{box} = [rectangle, minimum width=2cm, minimum height=1cm,text centered, draw=black, fill=blue!20]
  \tikzstyle{lbox} = [rectangle, minimum width=5cm, minimum height=1cm,text centered, draw=black, fill=blue!20]

 \node (SCP) [lbox, above of=sentiment] {Sentiment Contribution};
 \node (Tal) [box, above of=tally] {Tallying};
 \node (PE) [box, above of=pe] {Performance Evaluation};
 \tikzstyle{thinarrow} = [->,>=stealth]
 \draw [thinarrow] (SCP) -- (Tal);
 \draw [thinarrow] (Tal) -- (PE);
%
 
 \tikzstyle{arrow} = [thick,->,>=stealth]
 \node (User) [box, above of=SCP,node distance=2cm] {User};
 \node (Userbelow) [below of=User,node distance=.15cm]{};
 \node (Userbottom) [right of=Userbelow,node distance=.9cm]{};
 \node (Usertop) [above of=Userbottom,node distance=.3cm]{};
 \draw [arrow] (Tal) |- (Userbottom);
 \draw [arrow] (PE) |- (Usertop);
 \draw [arrow] (User) -- (SCP);
 
 \draw[-] (5.75,0) node[anchor=north] {$\Delta T_0$};
 \draw[-] (8.25,0) node[anchor=north] {$\Delta T_1$};
 \draw[-] (3,2) node {stake};
 \draw[-] (6.65,2) node {\emph{(partially)} return stake};
 \draw[-] (10.65,2) node {reward};
 \end{tikzpicture}
\end{center}
\caption{Overview of the Sentiment Protocol \emph{(with penalties)}.}\label{FigOverview}
\end{figure}
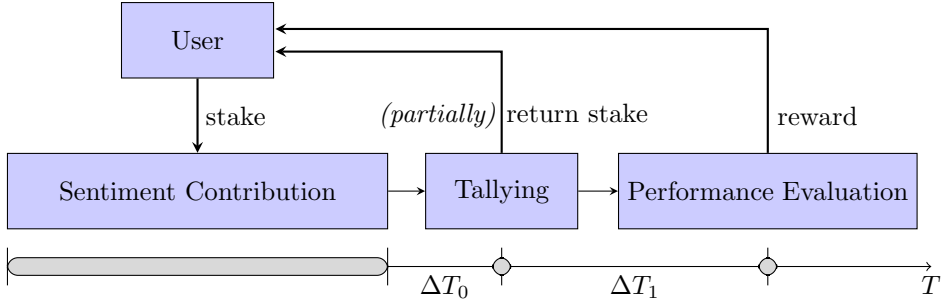

The Sentiment Protocol (cf. \Cref{FigOverview}) has three main components:
\begin{enumerate}
 \item Sentiment Contribution Period,
 \item Tallying and
 \item Performance Evaluation.
\end{enumerate}
During the Sentiment Contribution Period, users (sentiment providers) can submit their sentiment to the protocol. At the end of the Sentiment Contribution Period, and after a cool-down period $\Delta T_0$ has lapsed (which may be $0$), votes are tallied. The polling results can either be used for purely informational purposes or, in the use case of decentralized governance (cf. \Cref{UseGov}), directly trigger the policy that was voted on. After another cool-down period $\Delta T_1$ (which may again be $0$), users receive a (possibly negative) performance based reward.

\vspace*{.3cm}

\subsection*{Setting up a poll}
%
To set up a poll, the creator needs to provide the following:
\begin{enumerate}
 \item {\bf Topic and set of possible outcomes} $\rightarrow$ \Cref{Topic} 
 \item {\bf Staking parameters} $\rightarrow$ \Cref{Staking} 
 \item {\bf Information on usage of results}  $\rightarrow$ \Cref{Tallying}  
 \item {\bf Performance evaluation parameters} $\rightarrow$ \Cref{PerformanceEvaluation} 
\end{enumerate}
\subsection{Topic and outcome set}\label{Topic}
The creator of the poll has complete freedom over the choice of topic and also the set of possible outcomes $\O $. Outcomes can be discrete, such as multiple choice, or continuous, such as real numbers. The creator also has the option of using a public key encryption scheme for the submission of sentiment. By doing so, third parties are prevented from obtaining knowledge of the already submitted sentiment.  

\begin{Example}[Discrete Outcome set $\O $] \label{ExDisOut}
The poll wants to predict the outcome of the 2020 US presidential election. The outcome set is $\O =\{R,D\}$, where $R$ is the republican and $D$ the democratic candidate.
\end{Example}

\begin{Example}[Continuous Outcome Set $\O $]\label{ExContSet}
The poll aims to predict the performance of the TESLA stock from \dat{2017}{11}{01} to \dat{2018}{05}{01}. The outcome set is $\O =[0,\infty)$ and $\O \ni o=\frac{p(2018-05-01)}{p(2017-11-01)}$, where $p$ is the price of TESLA at a given date.
\end{Example}

\subsection{Providing sentiment by staking}\label{Staking}
To provide sentiment, each participant needs to stake a corresponding amount of assets. For simplicity, we focus on the typical use case where the user needs to stake an ERC-20 token into the polling smart contract. The set of possible sentiments $ \S \subset \mathcal{P}(\O )$ is a subset of the powerset of the outcome set $\O $.\footnote{Recall, that the powerset $\mathcal P(\O )$ of a set $\O $ is the set of all subsets of $\O $.}\footnote{For the technical reader: if $\O $ is continuous we are using the Borel sigma algebra $\sigma(\mathcal{O})$ instead of $\mathcal P(\O )$.}

During the sentiment contribution period any participant that holds the token required for staking can submit his sentiment $s\in \S $. The weight of the vote is directly proportional to the number of tokens submitted, i.e. $1$ \tok\ = $1$ \sen. The setup of the staking phase requires: 
\begin{itemize}
 \item {\bf Token type} 
 \item {\bf Sentiment Contribution Period} 
 \item {\bf Limits on total submissions} 
\end{itemize}

Having to commit tokens with each choice limits the amount of sentiment each participant can submit. Furthermore, in cases where the performance evaluation (cf. \Cref{PerformanceEvaluation}) can lead to penalties, the commitment of tokens exposes the provider to a financial risk. It is important to note that each staked token has equal weight in the sentiment contribution process. In particular, the rewards and/or penalties are directly proportional to the amount of tokens staked. This choice is due to the pseudonoymous nature of blockchain, where it is futile to limit sentiment per address. However, each pollster is free to use a custom staking token. Such a custom token could for instance be issued by the polling company and may have restrictions on transferability. Even in cases where no penalty is possible, obtaining a large amount of staking tokens puts the buyer at the volatility risk of the token price. Moreover, even if contributors are willing to accept the volatility risk, as a major stakeholder, they would have little interest in a behaviour that would undermine the polling process.

Once the Sentiment Contribution Period has ended and a time $\Delta T_0$ (which may be $0$) has lapsed, the votes are tallied. If the votes were encrypted, the pollster will have to use his private key to encrypt the submissions. If the performance evaluation does not involve penalties, the stake is returned to the sentiment provider. If penalties are possible, the stake minus the maximal possible penalty is returned. 

\begin{Example}[Staking without penalties] \label{ExStaking}
Let us return to the previous example of the 2020 presidential election and assume that no penalties are possible. We set $\Delta T_0=24h$, $\S  = \O =\{R,D\} $ and  choose the following staking parameters:\footnote{For the technical reader: we identify $\{\{R\},\{D\}\} \cong \{R,D\}$.}
\begin{itemize}
 \item Type: ETH,
 \item Sentiment Contribution Period: \dat{2017}{12}{01} to \dat{2017}{12}{10},
 \item Limits on total submissions: Minimum $1000$ ETH ($1000$ \sen) and maximum $10{,}000$ ETH ($10{,}000$ \sen).
\end{itemize}
Thus, a sentiment provider who wants to provide $100$ \sen\ for the democratic candidate $D$ needs to submit $100$ ETH together with his (possibly encrypted) choice $D$ to the polling smart contract.\footnote{For the technical reader: the user needs to encrypt the value with a nonce and also submit the encrypted nonce.} Since no penalties are possible, $100$ ETH are returned to the sentiment provider on \dat{2017}{12}{11}.
\end{Example}

\subsection{Usage of results}\label{Tallying}
Apart from simply aggregating information on behalf of the pollster, one could also link the execution of certain events to the results of the tally. For instance, one could ask individuals to estimate their energy consumption and, if it is below a certain threshold, a power plant could be idled. If individuals report inaccurate information they could be penalized by either not earning rewards or by loosing some of their stake. Another -- often controversial -- use case are decentralized autonomous organizations \cite{Aragon}. In this case, the polling results could be used to autonomously implement policy changes within the organization (cf. \Cref{UseGov}).

\subsection{Performance Evaluation} \label{PerformanceEvaluation}
The payment of performance based rewards is a key feature of the Sentiment Protocol. The distribution of rewards/penalties is determined by the performance evaluation function provided by the poll creator. 

\begin{Def}[Performance evaluation function \pef]\label{DefPEF}
The performance evaluation function $\pefm:\O  \times \S \rightarrow \mathbb [-1,\infty)$, depends on the outcome $o\in \O $ and the submitted sentiment $s\in \S$ and specifies the reward/penalty per submitted token.\footnote{The technical reader may have noticed that the triple $(\O,\S,\pefm)$ is reminiscent of a \href{https://en.wikipedia.org/wiki/Probability_space}{probability space}. Of course, this is not a coincide.}
\end{Def}

The supremum\footnote{If $|\O |< \infty$, $\sup \pefm=\max \pefm$.} of the performance evaluation function determines the size of the reward pool that needs to be provided to set up the poll. It is given by
\begin{equation*}
 (\mathrm{reward\ pool})= \sup \pefm \cdot (\mathrm{limit\ on\ total\ submissions}),
\end{equation*}
where all values are in units of tokens. If $\pefm\geq0$ (as a function), the poll does not involve penalties and thus the stake is returned immediately after tallying. If $\pefm$ takes negative values, $(1+\inf \pefm)\cdot T$ tokens are returned after tallying, where $T$ is the number of submitted tokens.

Once the performance evaluation time is reached ($\Delta T_1$ after tallying), the rewards are determined and payed out to the sentiment providers. Any remaining balance in the reward pool is returned to the pollster. See also \Cref{ApMultiPE} for an extension of the protocol that allows multiple performance evaluations. 

\begin{Example}[Constant \pef]\label{ExConstant}
 In this trivial case $\pefm\equiv c$ for some positive constant $c\in \mathbb R^+$. Thus, the reward is independent of both the provided sentiment and the outcome. Assuming the same staking parameters as in Example \ref{ExStaking}, the reward pool is $10{,}000 \cdot c$ and, since the reward is independent of the outcome or sentiment, there is no need to wait ($\Delta T_1=0$) and all sentiment providers receive $(1+c)T$ tokens on \dat{2017}{12}{11}, where $T$ is the number of submitted tokens. This trivial case corresponds to the current practice of most polling companies.
\end{Example}

\begin{Example}[Positive \pef for discrete $\O $] \label{ExElection2020}
We return to Example \ref{ExStaking}. To incentivize sentiment providers to make good predictions we define 
\begin{equation*}
 \pefm(D,D)=\pefm(R,R)=c \mbox{ and } \pefm(D,R)=\pefm(R,D)=0. 
\end{equation*}
Thus, only people that voted for the winning candidate get rewarded. Since the results of the elections will not be finalized until mid November 2020, $\Delta T= 1071$ days. As in Example \ref{ExConstant}, the reward pool is $10{,}000 \cdot c$.
\end{Example}

\begin{Example}[\pef for continuous $\O $]\label{ExStock}
 We return to Example \ref{ExContSet} and use the same staking parameters as in Example \ref{ExStaking}. The sentiment set is $\S=\{\mathrm{buy},\mathrm{sell}\}$ and we wish to choose \pef such that providers of buy ratings get rewarded for a positive development of the TESLA stock price ($o>1$) and penalised for a negative development.\footnote{We can identify the \emph{sell} rating with $[0,1)\in \mathcal P(\O)$ and the \emph{buy} rating with $(1,\infty)\in \mathcal P(\O)$.} Similarly, providers of sell ratings shall get rewarded for a negative development of the TESLA stock price ($o<1$) and penalised for a positive development. In the interest of fairness, if the stock price e.g. doubles ($o=2$), the buy rater should get the same reward as a sell rater if the stock price halves ($o=1/2$). A natural choice for \pef that satisfies these requirements and is also bounded is 
 \begin{equation*}
  \pefm(o,s)=\frac{2 c}{\pi}\cdot \operatorname{sgn}(o-1)\cdot {\arctan\left(\max\left\{o-1,o^{-1}-1\right\}\right)}\cdot g(s),
 \end{equation*}
where $g(\mathrm{buy})=1$ and $g(\mathrm{sell})=-1$. Note that $\arctan$ has the appealing properties of being approximately linear around $0$, strictly monotone increasing but still bounded by $\pi/2$. Thus for small price changes
\begin{equation*}
 \pefm(o,\mathrm{buy})\approx \begin{cases}
                             c(o-1), &o\geq1 \\
                             -c(o^{-1}-1), &o<1 
                            \end{cases}
\end{equation*}
and similarly $\pefm(o,\mathrm{sell})\approx -c(o-1)$ for $o\geq1 $ and $\pefm(o,\mathrm{sell})\approx c(o^{-1}-1)$ for $o<1$. Owing to monotonicity, higher (sentiment aligned) performance results in higher rewards. As in the previous examples, the reward pool is $10{,}000 \cdot c$. The maximal possible penalty is $-c$ and thus only $(1-c)T$ tokens are returned after tallying. 
\end{Example}

\section{Incentives and vulnerability analysis}\label{SecInc}
In this section, we analyse the incentives for pollsters and sentiment providers to use the Sentiment Protocol. We distinguish between public polls and polls amongst experts. We also explain its advantages over current (centralized) polling mechanisms and discuss possible vulnerabilities. 

\subsection{Benefits of using blockchain technology}
Firstly, it is important to note that opinion polling is not a truly decentralized mechanism and in particular the Sentiment Protocol does not claim to be fully decentralized. Nonetheless, using blockchain technology offers important advantages over classical, completely centralized, solutions. For the Sentiment Protocol, the two most important properties of blockchain technology are:
\begin{enumerate}
 \item the ability to easily transact and store value without the need of trusting a third party and \label{trussless}
 \item the ability to immutably store data. \label{store}
\end{enumerate}
The Sentiment Protocol uses \ref{trussless} for both the staking and the rewards/penalties. In a centralized system, users are hesitant to commit a significant value for period of time ($\geq\Delta T_0$) for the chance of receiving a comparatively small reward. Furthermore, by additionally using \ref{store}, the submitted sentiment is immutably stored and the user can easily prove that he is entitled to receive a reward - which may be far ahead in the future (cf. Example \ref{ExElection2020} and \Cref{ApMultiPE}).

\subsection{Public polls}\label{SecPublic}
The goal of a company conducting public polls is to receive reliable sentiment from a large number of users. The users on the other hand wish to be rewarded for providing accurate information. In the following we describe a possible setup of the Sentiment Protocol to align their interests.

To ensure the diversity of sentiment providers, the polling company issues their own token called POLL which can be arbitrarily transferred to and from polling smart contracts but cannot be transferred between users. However, all earned rewards are freely transferable. Further, we assume that at most times there are several polls to chose from and that each user has a starting balance of $100$ POLL. The performance evaluation function \pef is chosen positive (i.e. no penalties) and the cool down period $\Delta T_0$ is one week. Due to this long cool-down period, POLL holders are incentivized to participate in polls where they feel most confident in predicting the result. Unless they have absolute certainty that their prediction is correct, they are also incentivized to diversify their tokens on several polls. If however, they (think that they) are absolutely certain about the outcome, they may be inclined to stake all their POLL tokens on a single event. Of course, such behaviour is also in the interest of the polling company.

\subsection{Experts opinions}\label{SecExpert}
The goal of the polling company is to receive high quality predictions. However, in contrast to public polls, the users (experts) may wish to demonstrate their commitment to a forecast by putting their own capital at risk and thus earn credibility. 

Since the users have an interest in demonstrating credibility, the polling company chooses a performance evaluation function with penalties. Since each submission of sentiment may result in a loss of value, the company can simply choose ETH as a staking token. A sentiment provider that feels more certain about a prediction will choose to stake more tokens than a provider who is not as certain. As in the previous case, this behaviour is in the interest of the polling company.

\subsection{Limitations and possible vulnerabilities}
An important limitation of blockchains is their inability to source external information. Thus, in order to determine rewards, users needs to rely on oracles. This problem is not specific to the Sentiment Protocol but poses a general problem in developing truly decentralized applications. Another important limitation of smart contract platforms is their high cost for computations. Thus, the cost for determining performance based rewards may be significant. 

If these general limitations of blockchains can be overcome (or are irrelevant as in Example \ref{ExConstant}) and if the incentive structure in the Sentiment Protocol is set up correctly, it is impossible for the pollster or the user to illegitimately receive funds.

Owing to currently having these limitations, the polling company may act in a much more centralized manner and may chose to conduct the rewards calculations off-chain with data provided by regular (centralized) data-feeds. Of course, this special role allows the pollster to illegitimately keep rewards. However, due to the public availability of the provided sentiment and the performance evaluation function, cheating by the pollster can easily be proven and the polling company will quickly loose credibility. 
Thus, cheating by the company becomes unprofitable in the long run.

To consider possible cheating from the users perspective it is convenient to distinguish between performance evaluation function with and without penalties. If no penalties are possible then a user has no risk of loss. However, even if the pollster does not use a custom token as detailed in \Cref{SecPublic}, the commitment of huge amounts of tokens is on the one hand very costly and on the other hand only profitable if the predictions are right in the end. Furthermore, if the token is native to the polling company, amassing large amounts makes the user a major stakeholder of the company and thus any misbehaviour counterproductive. 

If penalties are possible and if potential rewards/penalties are sufficiently large, it may be advisable to choose a `zero-sum' performance evaluation function, i.e. there should not exist a combination of sentiments and staking choices such that the sentiment provider earns a profit independent of the outcome. Of course, as in the case without penalties, it is at the discretion of the pollster to allow this possibility. 
%
%
%
%

\section{Use cases}\label{SecUse}
In this section we discuss two specific use cases. We will not discuss the two cases in full details but rather assume familiarity with \Cref{SecSentProtocols}.
\subsection{Governance}\label{UseGov}
Any token that is able to access smart contract functionality (e.g. ERC-20 tokens) can use the Sentiment Protocol for decentralized governance. Since policy changes are not objectively verifiable as right or wrong, it is natural to chose a constant performance evaluation function $\pefm\equiv c$ (cf. Example \ref{ExConstant}). Thus, each token holder that participates in the governance process gets a fixed reward for his engagement. The required reward pool can be provided by the token issuing company. To ensure an indefinite supply of reward pools $R_i$ for the $i$th voting event, the company can e.g. use a distribution according to a geometric series. That is, if the company plans on using $100{,}000$ tokens for incentivizing governance participation (via the Sentiment Protocol), they can set $R_{i+1}=x\cdot R_{i}$, where $x<1$ and\footnote{Indeed, $\sum_{i=1}^\infty R_i=100{,}000\, (1-x) \sum_{i=0}^\infty x^{i}=100{,}000\, (1-x) \frac{1}{1-x}=100{,}000$.}
\begin{equation}\label{EqReward}
 R_1=100{,}000 \cdot (1-x).
\end{equation} 
Since the performance evaluation function is independent of the outcome, neither oracles nor complex computations are necessary and thus the (governance) Sentiment Protocol can be entirely implemented on-chain. In particular, this allows running decentralized autonomous organization, where user are allowed to vote on policy changes and decisions are automatically executed according to the voting results.  
%

\subsection{Community driven rating agency}
In this scenario, the polling company wishes to regularly receive sentiment on the performance expectations of a wide range of stocks. To ensure broad reach, the polling company issues $1{,}000{,}000$ custom tokens of which $900{,}000$ are distributed to the public and $100{,}000$ are held by the company to fund reward pools. The transferability of originally distributed tokens between users is limited (e.g. only $10\%$ per quarter). However, earned rewards can be transferred freely. To ensure the indefinite supply of rewards pools, the company chooses $x=0.99=99\%$ (cf. \eqref{EqReward}) and thus the reward pool for the first sentiment round is $R_1=1000$ tokens and each subsequent pool is $99\%$ of the size of the previous pool. In each round the users are asked to provide their sentiment on the performance of $10$ different stocks within the next three months ($\Delta T_1=3$ months) by choosing between $\Uparrow$, $\Leftrightarrow$ and $\Downarrow$ for each stock. 

\begin{figure}[h]
\begin{center}
 
\begin{tikzpicture}
\put(3,100){\textcolor{red}{\LARGE$\Downarrow$}}
\put(52,98){\textcolor{blue}{\LARGE$\Leftrightarrow$}}
\put(150,100){\textcolor{green}{\LARGE$\Uparrow$}}

    \begin{axis}[
        domain=0.5:2,
        xmin=0.4545454, xmax=2.199,
        ymin=0, ymax=1,
        samples=400,
  axis y line=left,
axis x line=middle,
axis equal,
every axis x label/.style={at={(current axis.right of origin)},anchor=north},
minor tick num=1,
xlabel=$o$,ylabel=$\pefm/c$
    ]
        \addplot+[mark=none,domain=1:2,color=green] {x-1};
        \addplot+[mark=none,domain=2:2.5,color=green] {1};
        \addplot+[mark=none,domain=.5:1,color=red] {1/x-1};
        \addplot+[mark=none,domain=.4:.5,color=red] {1};
        \addplot+[mark=none,domain=.909090909:1,color=blue] {5*(1.1-1/x)};
        \addplot+[mark=none,domain=1:1.1,color=blue,dash pattern=on 40pt off 0pt] {5*(1.1-x)};
    \end{axis}
\end{tikzpicture}
\end{center}
\caption{The performance evaluation function $\pefm$ for the three sentiments $\Downarrow$, $\Leftrightarrow$ and $\Uparrow$. The red curve indicates the function $o \mapsto \pefm(o,\Downarrow)$, the blue curve the function $o \mapsto \pefm(o,\Leftrightarrow)$ and the green curve the function $o \mapsto \pefm(o,\Uparrow)$. Lines not shown correspond to $\pefm=0$.}\label{FigPerfStock}
\end{figure}
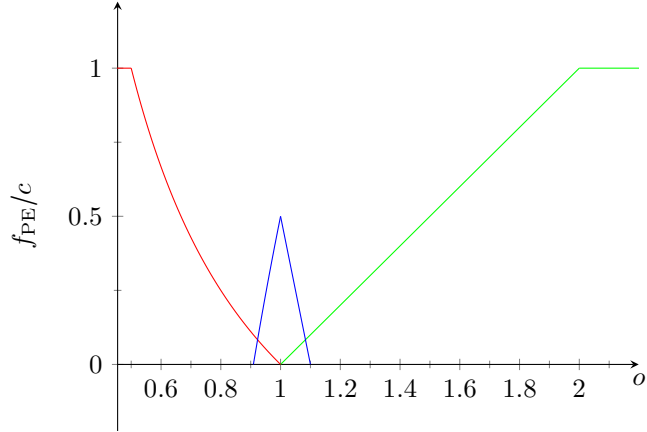

The performance evaluation function is defined by $\pefm(o,\Uparrow)=c\min\{1,\max\{0,o-1\}\}$, $\pefm(o,\Downarrow)=c\min\{1,\max\{0,1/o-1\}\}$ and 
\begin{equation*}
\pefm(o,\Leftrightarrow)=\begin{cases}
  5c(1.1-1/o) & o\in[0.\overline{90},1], \\
  5c(1.1-o) & o\in[1,1.1], \\
  0 &\mathrm{otherwise}.
  \end{cases}
\end{equation*}
Thus, a $\Uparrow$-rater gets rewarded only for positive stock performance and his reward is linear in the performance with a cap at $100\%$. A $\Leftrightarrow$-rater receives a maximal payout if the stock price does not change and does not get rewarded if the performance is outside the interval $(0.\overline{90},1.1)$. See \Cref{FigPerfStock} for a plot of this function. We note that the performance evaluation function only takes non-negative values and thus there is no risk of loss for participants. In particular, participation in the poll does not involve gambling.

Since the reward pool is $1000$ tokens and $900{,}000$ tokens are held by the sentiment providers we set $c=1000/(10\cdot 900{,}000) =0.000\overline1$. 
By allowing the rating of $10$ stocks simultaneously, users are incentivized to prioritize stocks (by staking more tokens) that they feel most confident in making a good prediction.  

The obtained sentiment data is a valuable asset to the company and could for instance be sold to third parties or used to create a community driven portfolio. In the latter case, the community itself would have the chance to invest in such a portfolio and thus double their incentive to participate in the rating process.

\section{Conclusions}
In this paper, we have shown how the Sentiment Protocol can leverage blockchain technology to align the incentives of pollsters and sentiment contributors. We discussed several use cases including decentralized governance and prediction markets. By introducing a fixed performance evaluation function we are able to reward predictions finely graduated. We have shown how such an approach offers a clear advantage over classical polling solutions, where rewards are either non-existing or small and independent of the quality of the contributed sentiment, and also over classical prediction markets, which are considered illegal in most jurisdictions.

\section*{Acknowledgments}
The research of A. M. leading to these results has received funding from the German Academic Exchange Service and the University of California, Berkeley. The author also expresses his gratitude to J. Benassaya for providing inspiration to this work and many helpful discussions.

\appendix
\section{Sentiment protocol with multiple performance evaluations}\label{ApMultiPE}
\begin{figure}[ht]
\begin{center}
 
 \begin{tikzpicture}


   \draw[->] (0,0) -- (12.15,0) node[anchor=north east] {$\Delta T_3$};
 
 \draw (0,.25) -- (0,-.25);
 \draw (5,.25) -- (5,-.25);
 \draw (6.5,.25) -- (6.5,-.25);
 \draw (9,.25) -- (9,-.25);
 \draw (11,.25) -- (11,-.25);
 
  \tikzstyle{period} = [rectangle, rounded corners, minimum width=5cm, minimum height=.1cm,text centered, draw=black, fill=gray!30]
 \tikzstyle{point} = [rectangle, rounded corners, minimum width=.1cm, minimum height=.1cm,text centered, draw=black, fill=gray!30]
 
 \node (start) {};
 \node (sentiment) [period, right of=start,node distance=2.5cm] {};
 \node (tally) [point, right of=start,node distance=6.5cm] {};
 \node (pe1) [point, right of=start,node distance=9cm] {};
 \node (pe2) [point, right of=start,node distance=11cm] {};

 \tikzstyle{box} = [rectangle, minimum width=1.5cm, minimum height=1cm,text centered, draw=black, fill=blue!20]
  \tikzstyle{ubox} = [rectangle, minimum width=2cm, minimum height=1cm,text centered, draw=black, fill=blue!20]
  \tikzstyle{lbox} = [rectangle, minimum width=5cm, minimum height=1cm,text centered, draw=black, fill=blue!20]
  \tikzstyle{tbox} = [rectangle, minimum width=2cm, minimum height=1cm,text centered, draw=black, fill=blue!20]

 \node (SCP) [lbox, above of=sentiment] {Sentiment Contribution};
 \node (Tal) [tbox, above of=tally] {Tallying};
 \node (PE1) [box, above of=pe1] {1. PE};
 \node (PE2) [box, above of=pe2] {2. PE};
 \node (PE3) [right of=PE2,node distance=1.2cm] {};

 \tikzstyle{thinarrow} = [->,>=stealth]
 \draw [thinarrow] (SCP) -- (Tal);
 \draw [thinarrow] (Tal) -- (PE1);
 \draw [thinarrow] (PE1) -- (PE2);
 \draw [thinarrow] (PE2) -- (PE3);
%
 
 \tikzstyle{arrow} = [thick,->,>=stealth]
 \node (User) [ubox, above of=SCP,node distance=2cm] {User};
 \node (Userbelow) [below of=User,node distance=.3cm]{};
 \node (Userbottom) [right of=Userbelow,node distance=.9cm]{};
 \node (Usertop1) [above of=Userbottom,node distance=.3cm]{};
 \node (Usertop2) [above of=Userbottom,node distance=.6cm]{};
 \draw [arrow] (Tal) |- (Userbottom);
 \draw [arrow] (PE1) |- (Usertop1);
 \draw [arrow] (PE2) |- (Usertop2);
 \draw [arrow] (User) -- (SCP);
 
 \draw[-] (5.75,0) node[anchor=north] {$\Delta T_0$};
 \draw[-] (7.85,0) node[anchor=north] {$\Delta T_1$};
  \draw[-] (10.15,0) node[anchor=north] {$\Delta T_2$};
 \draw[-] (3,2) node {stake};
 \draw[-] (6.66,2) node {\emph{(partially)} return stake};
 \draw[-] (9.85,2) node {1. reward};
 \draw[-] (11.85,2) node {2. reward};

 \end{tikzpicture}
\end{center}

\caption{Overview of Sentiment Protocol \emph{(with penalties)} with long term rewards.}\label{FigMulti}
\end{figure}
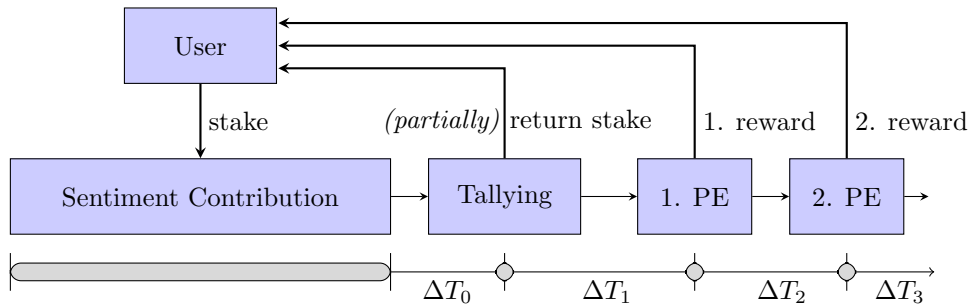

To incentivize users to provide sentiment on long term performance (e.g. stock prices), we extend the protocol to allow several performance evaluations (cf. \Cref{FigMulti}). Let us illustrate this by assuming that the performance is evaluated every three months, i.e. $\Delta T_i=3$ months. To this end, we can e.g. define the performance evaluation function ${\pefm}_i$ at the $i$th performance evaluation as
\begin{equation*}
 {\pefm}_i=2^{-i}\pefm,
\end{equation*}
where $\pefm$ is the original (single event) performance evaluation function. Since $\sum_{i=1}^{\infty}2^{-i}=1$, the reward pool does not need to be increased. Of course, the pollster is free to make different choices for ${\pefm}_i$. The only restriction is that the reward pool needs to be big enough to cover \emph{all} performance evaluations.

\end{document}